\documentclass[review,longtitle,number]{elsarticle}
\usepackage{graphicx}
\usepackage[square,numbers]{natbib}
\usepackage{amssymb}
\begin{document}
\journal{Physics Letters B}

\begin{frontmatter}
\title{
First underground results with NEWAGE-0.3a direction-sensitive dark matter detector}

\author{Kentaro~Miuchi$^a$},
\ead{miuchi@cr.scphys.kyoto-u.ac.jp}
\author{Hironobu~Nishimura$^a$}, 
\author{Kaori~Hattori$^a$}, 
\author{Naoki~Higashi$^a$}, 
\author{Chihiro~Ida$^a$},
\author{Satoshi~Iwaki$^a$}, 
\author{Shigeto~Kabuki$^a$}, 
\author{Hidetoshi~Kubo$^a$},
\author{Shunsuke~Kurosawa$^a$},
\author{Kiseki~Nakamura$^a$},
\author{Joseph~Parker$^a$},
\author{Tatsuya~Sawano$^a$},
\author{Michiaki~Takahashi$^a$}, 
\author{Toru~Tanimori$^a$},
\author{Kojiro~Taniue$^a$},
\author{Kazuki~Ueno$^a$}, 
\author{Hiroyuki~Sekiya$^b$},
\author{Atsushi~Takeda$^b$}
\author{Ken'ichi~Tsuchiya$^c$}, 
\author{Atsushi~Takada$^d$},
\address{$^a$ Cosmic-Ray Group, Department of Physics, Graduate School of Science, 
Kyoto University Kitashirakawa-oiwakecho, Sakyo-ku, Kyoto, 606-8502, Japan}
\address{$^{b}$  Kamioka Observatory, ICRR, The University of Tokyo
Higashi-Mozumi, Kamioka cho, Hida 506-1205 Japan}
\address{$^c$
National Research Institute of Police Science ¡¡6-3-1 Kashiwanoha, Kashiwa, Chiba, 277-0882, Japan 
}
\address{$^d$
Scientific Balloon Laboratory, ISAS, JAXA
Yoshinodai 3-1-1, Sagamihara, Kanagawa, 229-8510,Japan}

\begin{abstract}
A direction-sensitive dark matter search experiment at Kamioka 
underground laboratory with the NEWAGE-0.3a detector was performed.
The NEWAGE-0.3a detector is a gaseous micro-time-projection chamber 
filled with CF$_{4}$ gas at 152 Torr. 
The fiducial volume and target mass are
20 $\times$ 25 $\times$ 31 cm$^3$ and 0.0115 kg, respectively.
With an exposure of 0.524 kg$\cdot$days,
improved spin-dependent weakly interacting massive particle (WIMP)-proton 
cross section limits 
by a direction-sensitive method were achieved
including a new record of 5400 pb for 150 GeV/c${^2}$ WIMPs. 
We studied the remaining background and found that  
ambient $\gamma$-rays contributed about one-fifth of the remaining 
background and 
radioactive contaminants inside the gas chamber contributed the rest.

\end{abstract}


\end{frontmatter}

\section{Introduction}
\label{section:introduction}
The interest in the search for 
dark matter has been growing continuously 
since the late 1980s. 
Attention to this problem recently increased 
after the 
Wilkinson Microwave Anisotropy Probe all-sky observation \cite{ref:WMAP_2009}, 
Sloan  Digital Sky Survey 
large-scale structure measurements \cite{ref:SDSS2004a,ref:SDSS2004b},  
and supernovae data 
from two other experiments
(Supernova Cosmology Project \cite{ref:SN1998} 
and High-Z Supernovae Search \cite{ref:SN1999} ) 
together produced more precise cosmological parameter determinations.
Weakly interacting massive particles (WIMPs) are one of the 
strongest candidates for dark matter. 
WIMPs are searched for mainly in three ways: 
collider experiments \cite{ref:LHC}, 
indirect (astrophysical) experiments \cite{ref:Fermi,ref:HESS}, 
and direct searches in the laboratory.
The Large Hadron Collider (LHC) experiment is expected to 
discover or set a stringent limit on the properties of  
super-symmetric particles, 
which are a good candidate for WIMPs. Indirect experiments 
might detect some clues to the nature of WIMPs.
While these experiments would provide information about
the masses and cross sections, 
we still need direct searches to demonstrate that
these particles are the dark matter in the Universe. 

Many direct search experiments for WIMPs have been 
concluded \cite{ref:DAMA_annual2000,ref:XENON10,ref:NAIAD,ref:Tokyo_CaF2}, 
are currently being performed \cite{ref:CDMS2,ref:DAMA_LIBRA,ref:KIMS,ref:COUPP}, or are being planned.
The experiments with liquid noble-gas detectors \cite{ref:XENON10} 
and cryogenic detectors \cite{ref:CDMS2} 
will be scaled up to more
than one kilogram to explore the region predicted by 
Minimal Supersymmetric extension of the Standard
Model (MSSM) theory \cite{ref:SUSYNillesb} in the next decade. 
These massive detectors only measure the energy 
transferred to the nucleus through WIMP-nucleus
scattering, thus the most distinct signal of WIMPs is annual
modulation of the energy spectrum. Because the amplitude of the
annual modulation signals is very small (a few percent of the total event
of WIMP-nucleus scattering), the result of an 11-cycles' annual
modulation observation reported by the DAMA/LIBRA group 
\cite{ref:DAMA_annual2000,ref:DAMA_LIBRA}
is the only positive signature ever reported. Although other
groups have tried to confirm the DAMA/LIBRA results with
various types of detectors, no group has observed an annual 
modulation signal yet. 
Many groups are preparing larger-mass detectors for a
nuclear-model-independent study of dark matter.
Furthermore, detection methods other than the annual modulation signature are
also necessary for an astrophysical-model-independent study.

Another distinct signature of WIMPs is thought to result from
the relative motion of the solar system with respect to the 
galactic halo.
If we assume isotropic WIMP motion,
the peak WIMP flux should  come from the direction
of solar motion, which happens to point toward the constellation 
Cygnus. 
The recoil rate would then peak in the opposite direction, and 
this distribution would be a distinct signal. 
Several experimental and theoretical works on
the possibility of detecting this anisotropy, or the WIMP-wind, 
have been performed so far 
\cite{ref:Anne,ref:DAMA_aniso1,ref:DAMA_aniso2,ref:sekiya,ref:DRIFT_PRL1994,ref:NEWAGE_PLB2004,ref:DMTPC_APP2008,ref:NIT_NIM2007}.
Among these proposed methods, 
a gaseous detector is one of the most appropriate devices for detecting this 
WIMP-wind because nuclear recoil tracks
can in principle be detected with better angular resolutions than
by other detectors \cite{ref:TPCforDM1,ref:TPCforDM2,ref:TPCforDM3}.
The DRIFT group has pioneered studies of gaseous detectors 
for WIMP-wind detection for more than ten years with multi-wire 
proportional chambers \cite{ref:DRIFT_NIM2009,ref:DRIFT_APP2007}.
We proposed a new project,  NEw  generation  WIMP-search
with Advanced Gaseous tracking device Experiment (NEWAGE) 
\cite{ref:NEWAGE_PLB2004}, 
which has advantages over the DRIFT detectors
in the pitch of the detection sensors and a 
three-dimensional tracking scheme. 
After the construction of a prototype detector 
and the first dark matter search experiment in 
a surface laboratory, we installed our prototype detector in an underground 
laboratory and studied its performance precisely \cite{ref:NEWAGE_PLB2007,ref:NEWAGE_APP2009}. 
In this paper, we report the results of our first underground direction-sensitive dark matter search experiment.

\section{Detector}
\label{section:detector}
For this experiment we used the NEWAGE-0.3a detector, 
the first prototype 
of our $\rm(0.3 m)^3$-class 
gaseous time-projection-chamber(TPC) series.
The detector system and performance studies 
are described in our previous work
\cite{ref:NEWAGE_APP2009}, so we briefly summarize the 
essential properties 
closely related to this dark matter search experiment in this section.

\subsection{System}
\label{subsection:system}
\begin{figure}
\includegraphics[width=1.\linewidth]{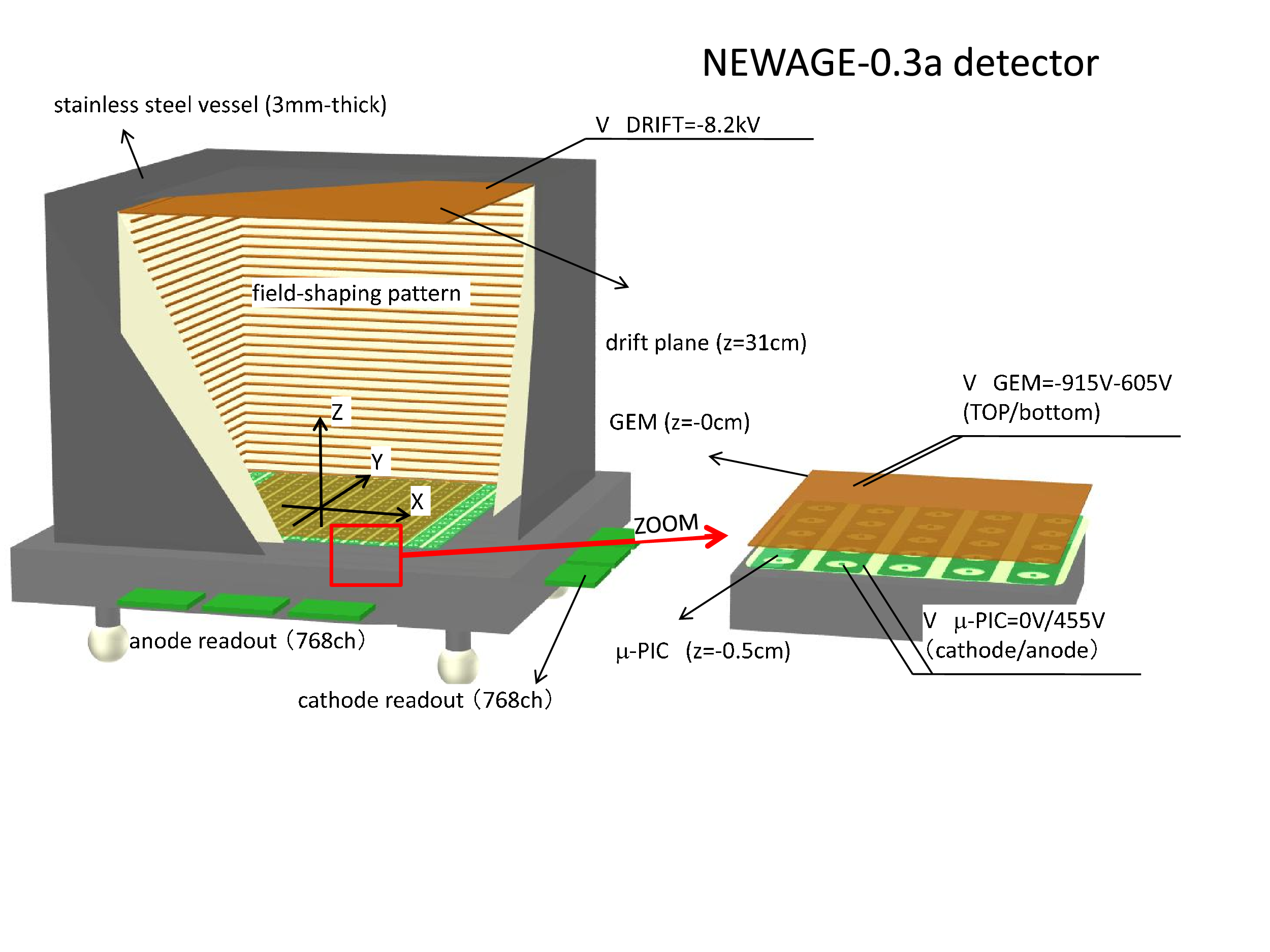}
\caption{\label{fig:TPC}
Schematic view of NEWAGE-0.3a detector.
The volume between the drift plane 
and the GEM is the detection volume, it is filled with 
CF$_4$ gas at 152 Torr.}
\end{figure}

A schematic view of the NEWAGE-0.3a detector is shown in FIG. \ref{fig:TPC}.
The NEWAGE-0.3a detector is a gaseous three-dimensional tracking detector 
read by a 30.7 $\times$ 30.7 cm$^2$ $\mu$-PIC (TOSHIBA/DNP,
SN060222-3). A $\mu$-PIC is a two-dimensional imaging device which has 
orthogonally-formed readout 
strips with a pitch of 400 $\mu$m \cite{ref:30uPIC}. 
Field-shaping patterns on fluoroplastic
circuit boards
form a detection volume above a gas electron multiplier (GEM) \cite{ref:GEM}. 
We used a GEM with an amplification area of 23 $\times$ 28 cm$^2$ 
(Scienergy Co. Ltd.) as an intermediate amplifier.
A fiducial volume of 20$\times$25$\times$31 cm$^3$
was defined in a detection volume of 23 $\times$ 28 $\times$ 31 cm$^3$. 
Among many potential candidates for the chamber gas, we used
CF$_4$, which has advantages for spin-dependent (SD)  WIMP detection.
CF$_4$ is also known to be a good TPC gas
because of its small diffusion coefficient, which is
an indispensable property for a good angular resolution detector.
We filled a stainless steel vessel 
with  CF$_4$ gas at 152 Torr.
The target mass in the effective volume was 0.0115kg.
A SORB-AC cartridge pump (SAES Getter MK5) was attached to the vessel
to absorb the out-going gas from the detector components.
Typical operation parameters, 
optimized to realize a stable
operation with a combined 
gas gain of 2400
($\mu$-PIC$\times$GEM = 300 $\times$8), are shown in Fig. \ref{fig:TPC}.

\subsection{Performance}
\label{subsection:performance}
We studied the detector performance in 
the underground laboratory. 
The measurements were carried out in the same 
manner as the dark matter runs in terms of the triggering 
and data acquisition systems.
One set of event data consists of track shape
information (digital hit points) and energy information (summed analog 
waveform).
We calibrated the energy
using the $\alpha$ particles generated in a $\rm^{10}$B(n,$\alpha$)$\rm^{7}$Li 
reaction (Q = 2.310 or 2.792 MeV).
We set a glass plate coated with a thin 0.6 $\rm \mu m$ $\rm{}^{10}B$ 
layer in the gas volume, and irradiated the detector with thermalized neutrons.
Alpha particles (5.6, 6.1, and 7.2 MeV) from decays of the
radon progeny were also used.
We thus used alpha-particle equivalent as the energy unit (keV$\alpha$.e) in this experiment.
(We took account of the ionization efficiency, 
or quenching factor, of fluorine and helium nuclei in the analysis 
to calculate the expected spectra for WIMPs.
We confirmed the linearity down in the dark matter energy range 
by the correlation of the track length and its energy.

We applied the following four event-selection criteria to the data.
These criteria are the same as those applied to data from
dark matter runs.
\begin{itemize}
\item N$\rm_{hit}$ selection:
We selected events that have at least three hit points. 
We needed this criterion to determine the track directions  
with a certain angular resolution.
\item Fiducial-volume selection: 
We selected events whose hit points were all in the fiducial volume.
This rejected the nuclear-track 
background events from the fluoroplastic walls of the detection volume.
\item Energy selection: We selected events whose energy is between 
100~keV$\alpha$.e.~\footnote{
100~keV$\alpha$.e. corresponds to 140 keV fluorine recoil, for reference.}
and 400~keV$\alpha$.e. 
The lower energy threshold is chosen so as to maintain 
a certain angular resolution, 
which is restricted by the length of the nuclear recoil. 
The higher one is set at the expected highest recoil energy of 
WIMPs at escape velocity. 
We refer to this as the DM energy range in the following discussion.
\item Nuclear-recoil selection: 
We select events with track lengths shorter than  
1 cm. The length limit rejects background $\gamma$-ray events.
\end{itemize}

We review the methods and results of detector performance measurements 
in the following paragraphs.
Details can be found in ref \cite{ref:NEWAGE_APP2009}.

\begin{itemize}
\item Energy resolution: The energy resolution was determined by  
three components: the electric noise term ($\rm\sigma_{noise}$), 
the gain non-uniformity term ($\rm \sigma_{non-uni})$,  
and statistics of the primary ion-electron pair term ($\rm \sigma_{sta}$).
$\rm\sigma_{noise}$ was measured to be 55$\%$,  
$\rm \sigma_{non-uni}$ was known to be 45$\%$ from the high-energy 
(6 MeV$\alpha$.e.) peaks,  and 
$\rm \sigma_{sta}$ was calculated to be 6$\%$.
Thus, the squared sum of these three terms gives an energy 
resolution of 70$\%$ (FWHM) at 100 keV$\alpha$.e. 
\item $\gamma$-ray detection efficiency: We irradiated the NEWAGE-0.3a 
with $\gamma$-rays from 
a $^{137}$Cs radioactive source. We compared the detection rate of 
electron tracks that passed event selection with 
a simulated rate;
the detection efficiency was found to be 
8.1$\times10^{-6}$  at 100 keV$\alpha$.e. 
Thus, the $\gamma$-ray rejection power was 99.9992$\%$ 
at 100 keV$\alpha$.e.
The energy dependence is shown in our previous work\cite{ref:NEWAGE_APP2009}.

\item Nuclear track absolute detection efficiency: 
We irradiated the NEWAGE-0.3a 
with neutrons from a $^{252}$Cf radioactive source. 
We compared the rate of 
detected nuclear tracks that passed event selection with 
the simulated rate. 
The nuclear track detection efficiency was found to be 
80$\%$ at 100 keV${\alpha.e.}$ 
The energy dependence is shown in our previous work\cite{ref:NEWAGE_APP2009}.
The measured energy dependence was taken into account in the 
analysis of dark matter run data.
\item  Direction-dependent nuclear track detection efficiency:
We irradiated the NEWAGE-0.3a 
with neutrons from a $^{252}$Cf radioactive source. 
We made isotropic scattering by placing the source at six positions 
and measured the direction-dependent nuclear track detection efficiency.



\item Nuclear track angular resolution: 
We irradiated the NEWAGE-0.3a 
with neutrons from a $^{252}$Cf radioactive source. 
We fitted the $|\cos \theta|$ distribution of recoil nuclear tracks 
with simulated ones smeared by various angular resolutions.
$\theta$ is the angle between the direction of the 
incident neutron and that of a detected nuclear track. 
We obtained an angular resolution of 55$^{\circ}$ (RMS) for 100 keV$\alpha$.e. 
nuclear tracks.
\end{itemize}

Typical results of performance measurement 
are listed in TABLE \ref{tab:detector}.

\begin{table}
\caption{\label{tab:detector}
Performance of NEWAGE-0.3a detector at energy threshold (100~$\rm keV{\alpha.e.}$)\cite{ref:NEWAGE_APP2009}.
}
\begin{tabular}{ll}
Parameter&Value\\
\hline
Energy resolution & 70\% (FWHM)\\
$\gamma$-ray detection efficiency & $\rm 8.1 \times 10^{-6}$ \\
Nuclear track detection efficiency&80\%\\
Nuclear track angular resolution& $\rm 55^\circ$(RMS) \\
\hline
\end{tabular}
\end{table}

\section{Measurements}
\label{section:measurements}
The first underground dark matter run with the NEWAGE-0.3a detector 
took place from September 11, 2008 until December 4, 2008 
in Laboratory B, Kamioka
Observatory (36$^{\circ}$25'N, 137$^{\circ}$18'E) 
located at 2700 m water-equivalent underground. 
The detector was set so
that the $\mu$-PIC plane was horizontal
and the X-axis was aligned in the direction of S87$^{\circ}$E.
Since the fast neutron flux in the underground laboratory was 
smaller than that in the surface laboratory 
by more than three orders of magnitude,  
the background in this first underground run was 
expected to be dominated by the internal background.
Therefore, we did not set any radiation shield. 
This dark matter run (N03aKa-Run5)
has three sub-runs, as listed in TABLE \ref{tab:subruns}.
We evacuated and refilled the vessel with new CF$_4$ gas at the beginning of 
each sub-run.

\begin{table}
\caption{\label{tab:subruns}
Summary of  NEWAGE-0.3a first underground dark matter run (N03aKa-Run5).
}
\begin{tabular}{llll}
Sub-Runs&Date&Live Time [days]&Exposure [kg$\cdot$days]\\
\hline
Run5-1&Sep. 11th -- Oct. 1st,  2008&17.81&0.204\\
Run5-2&Oct. 2nd  -- Nov. 11th, 2008&10.01&0.115\\
Run5-3&Nov. 13th -- Dec. 4th,  2008&17.90&0.205\\
\hline
\multicolumn{3}{r}{Total exposure}&0.524\\\hline

\end{tabular}
\end{table}

\begin{figure}
\includegraphics[width=1.\linewidth]{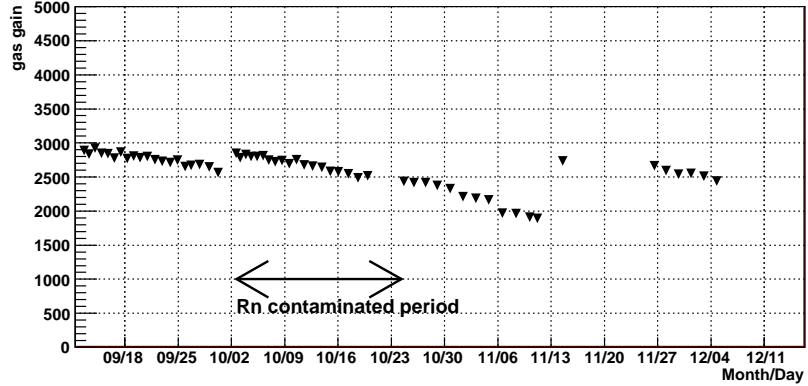}
\caption{\label{fig:gain}
Monitored gas gain during Run5.}
\end{figure}

\begin{figure}
\includegraphics[width=1.\linewidth]{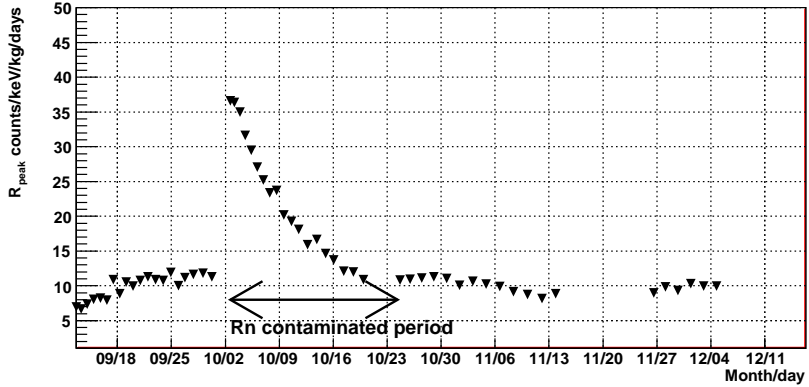}
\caption{\label{fig:radonrate}
Monitored radon progeny count rate during Run5.}
\end{figure}

We monitored the gas gain and the radioactive 
radon ($^{220}$Rn and $^{222}$Rn) 
contamination in the gas using high-energy ($\sim$6 MeV$\alpha$.e.) events.
The gas gain was monitored by the 
positions of the high-energy radon peaks.  
The monitored gas gains were used to correct the energy calibration.
The radioactive radon rate was monitored by the count rate of high-energy 
peaks so as to monitor part of the internal radioactivity.
The monitored gas gain and radioactive radon rate are shown 
in FIG. \ref{fig:gain} and FIG. \ref{fig:radonrate}, respectively. 
We found that we unintentionally 
contaminated the target gas with radioactive radon gas 
by using a vacuum tube and turbo molecular pump exposed to mine air  
during the gas replacement procedure at the beginning of Run5-2.
We therefore did not use this period for further analysis
(indicated in FIG. \ref{fig:gain} and \ref{fig:radonrate}).
Live times and exposures excluding this radon-contaminated period 
are shown in the run summary (TABLE \ref{tab:subruns}). 
A total exposure of 0.524 kg$\cdot$days was accumulated in about three months' 
measurement.

\section{Results}
\label{section:results}

\subsection{Measured Data}
\begin{figure}
\includegraphics[width=1.\linewidth]{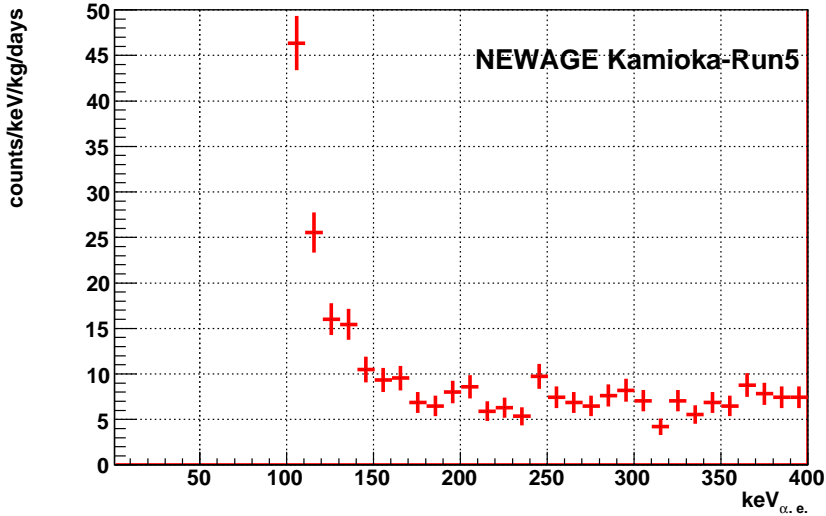}
\caption{\label{fig:spectrum}
Measured energy spectrum. The total exposure was 0.524 kg$\cdot$days.}
\end{figure}

\begin{figure}
\includegraphics[width=1.\linewidth]{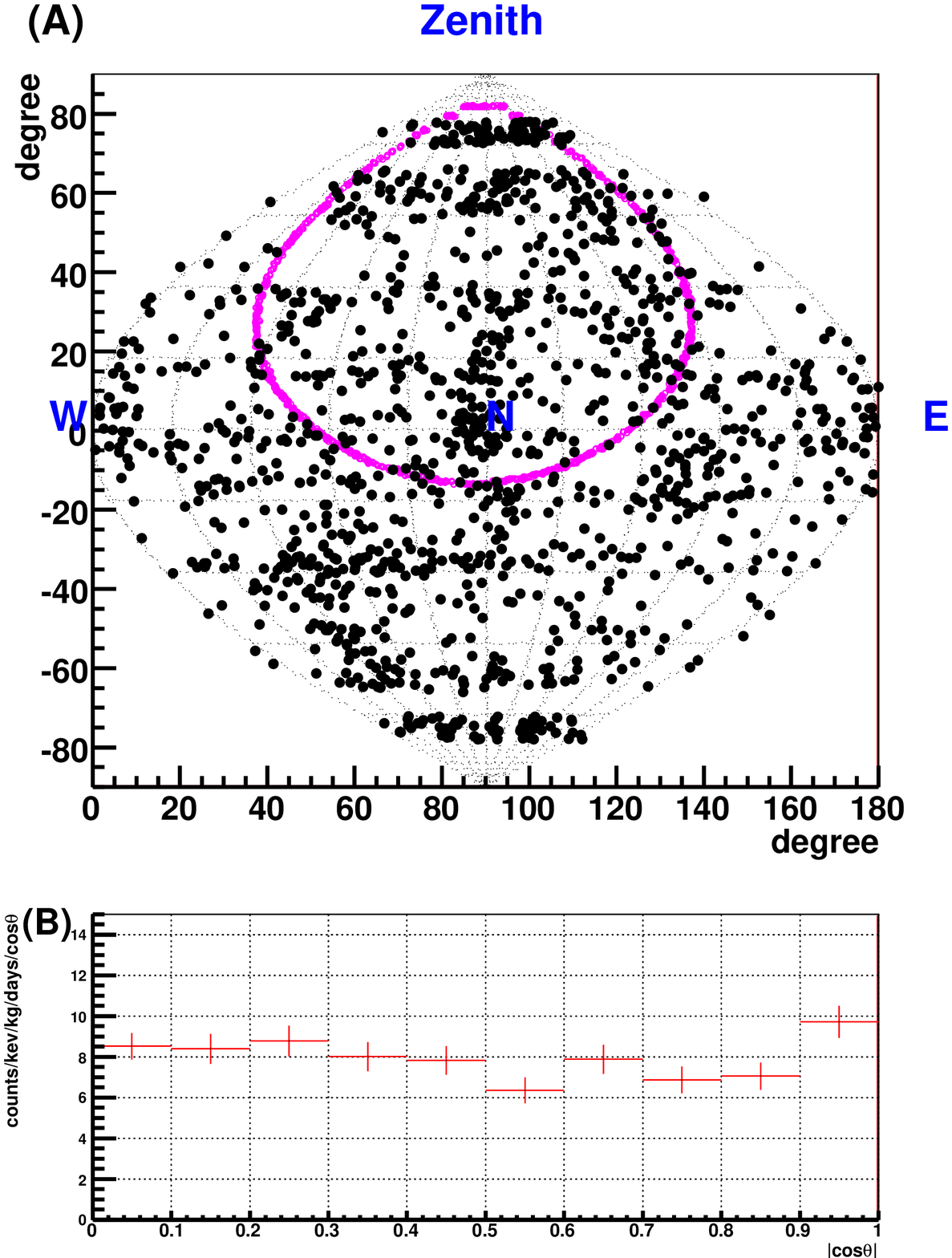}
\caption{\label{fig:skymap} 
Obtained directions of the nuclear tracks (A) and $|\cos \theta|$ 
distribution(B). 
Black markers in (A) indicate directions of the 1244 nuclear track events.
The direction toward Cygnus from which the WIMP-wind 
is expected at each event time is indicated by purple markers in (A).}
\end{figure}

We applied the data selection criteria described in Section
\ref{subsection:performance}
to the entire 0.524 kg$\cdot$days of 
Run-5 dark matter data. 
1244 nuclear tracks passed through the selections.
We corrected the count rate using the detection efficiency and 
obtained the energy spectrum shown in FIG.\ref{fig:spectrum}. 
The count rate at the energy threshold of 100keV$\rm \alpha.e$. was 
about 50 counts/keV/kg/days.
We also plotted the directions of all 1244 nuclear recoil tracks 
(shown by black markers in FIG.\ref{fig:skymap} (A)).
Because we did not detect
the sense of the nuclear tracks (track detection was axial-like  
and not vector-like), 
the map was restricted to half the
sky.
The southern part of the sky was folded into the northern half. 
We then performed a direction-sensitive analysis
assuming isotropic WIMP motion,
{\it i.e.}, WIMP-wind from the Cygnus direction.
The direction toward  Cygnus 
at each event time was also indicated by purple markers.
We calculated $\theta$, the angle between the recoil direction (black markers) 
and the corresponding WIMP-wind direction (purple markers) 
for each event, and made a $|\cos \theta|$ 
distribution which is shown in FIG. \ref{fig:skymap} (B).
The $|\cos \theta|$ distribution
is normalized with the energy range, the target mass, and live time, while 
it still contains detector responses such as the energy resolution, 
angular resolution, and detection efficiencies.

\subsection{Direction-Sensitive Analysis}
We derived direction-sensitive dark matter limits
by comparing the measured $|\cos \theta|$ distribution 
with those expected from WIMP-nucleus elastic scatterings. 
We used the astrophysical and nuclear parameters 
given in Table \ref{tab:params}.  
We used the same analysis procedure described in our 
previous work \cite{ref:NEWAGE_PLB2007} unless otherwise stated. 
We prepared the expected $|\cos \theta|$ distributions
for given WIMPs masses and energy bins. 
We considered the detector response, 
such as the energy resolution, 
angular resolution, and detection efficiencies, 
to make these expected $|\cos \theta|$ distributions.
Followings are the procedure for preparing this 
expected $|\cos \theta|$ distribution.

We followed ref. \cite{ref:LewinSmith} for the energy-spectrum calculation.
An ideal $|\cos \theta|$ distribution without detector responses can be 
calculated by equation \ref{eq:costheta}, where 
$R$ is the count rate,
$\theta$  is the recoil angle, $v_{\rm s}$  is the solar velocity 
with respect to the galaxy, $v_{\rm{min}}$  is the minimum velocity of
WIMPs that can give a recoil energy of $E_{\rm R}$, and $v_0$ is the
Maxwellian WIMP velocity dispersion 
\cite{ref:spergel,ref:DRIFT_PRL1994}.

\begin{equation}
\label{eq:costheta}
\frac{d^2R}{dE_{\rm R}d\cos \theta} \propto \exp\left[ \frac{ \left( v_{\rm s} \cos \theta -v_{\rm min}\right) ^2}{v^{2}_{0}} \right]
\end{equation}

\begin{figure}
\includegraphics[width=1.\linewidth]{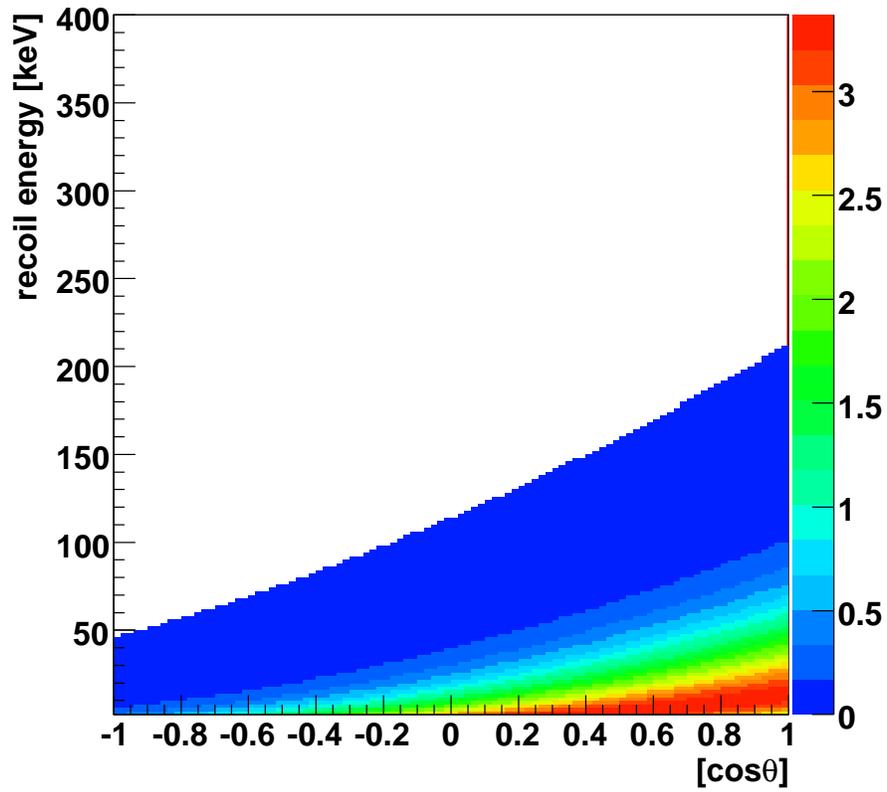}
\caption{\label{fig:E_cos} 
Expected two-dimensional event rate ``spectrum'' as a function of the 
recoil angle and recoil energy.
The target is fluorine nuclei.
A WIMP mass of 100 GeV/c$^2$ and a cross section of 1pb are assumed. 
$\theta$ is the recoil angle of
the nuclear recoil with respect to the WIMP-wind direction. 
Color contour shows event rate in the 
unit of [counts/keV/kg/days/cos$\theta$].
White areas are the parameter space where no event is expected.
}
\end{figure}

With the expected energy spectrum and the $|\cos \theta|$ distribution 
known from equation \ref{eq:costheta}, we made a 
two-dimensional event rate ``spectrum'' which is a function of the 
recoil angle and recoil energy(shown in FIG. \ref{fig:E_cos}).

The horizontal slice
of this two-dimensional spectrum 
is the ideal $|\cos \theta|$ distribution  
corresponding to the energy range of interest.
We took account of the detector response by the following steps.
\begin{itemize}
\item STEP 1: (For a given WIMP mass) 
We made an ideal two-dimensional event rate ``spectrum''
distribution like that shown in FIG. \ref{fig:E_cos}.

The input parameter is a WIMP mass.
The output is the two-dimensional event rate ``spectrum''.

\item STEP 2: (For a given WIMP-wind direction) 
We simulated a WIMP-nucleus scattering 
in the actual detector coordinates.
The recoil energy and angle $\theta_{\rm R}$ 
were generated according to the distribution calculated in STEP 1.
Here, we convert the recoil energy into alpha-equivalent energy, 
considering the 
ionization efficiencies, or quenching factors, of the fluorine and 
helium nuclei using the SRIM code\cite{ref:SRIM}.
The azimuth angle of the recoil,  $\phi_{\rm R}$, was randomly determined. 

The input parameter is a WIMP-wind direction. 
The output parameters are the direction and energy of a nuclear track 
in a detector.

\item STEP 3:  We calculated the detection efficiency 
as a product of the absolute detection efficiency which is 
a function of energy, and the relative efficiency which is a 
function of the track direction. 

Input parameters are the energy and direction of a nuclear track 
(result of STEP 2).
Output parameter is the detection efficiency.

\item STEP 4: We calculated the observable energy and direction 
by smearing the actual energy and direction according to the resolutions.

The input parameters are the actual energy and direction energy of a nuclear track
(result of STEP 2).
The output parameters are the observable direction and energy.

\item STEP 5: We calculated $|\cos \theta|$ using the observable 
track direction (result of STEP 4) and the 
WIMP-wind direction (input parameter of STEP 2).
We filled the $|\cos \theta|$ histogram of the 
energy bin of interest 
with the calculated  $|\cos \theta|$.
We used the efficiency calculated in STEP 3 as the weight for filling.

\item STEP 6:  We repeated STEPs 2 to 5 for many WIMP-wind 
directions to reproduce the 
actual direction distribution of the WIMP-wind 
during the observation time.

\item STEP 7:  We repeated STEPs  1 to 6 for the WIMP masses 
of interest.

\end{itemize}

In this way, we made $|\cos \theta|$ distributions 
expected from WIMP-nucleus elastic scatterings.

\subsection{Dark Matter Limits}
\begin{figure}
\includegraphics[width=1.\linewidth]{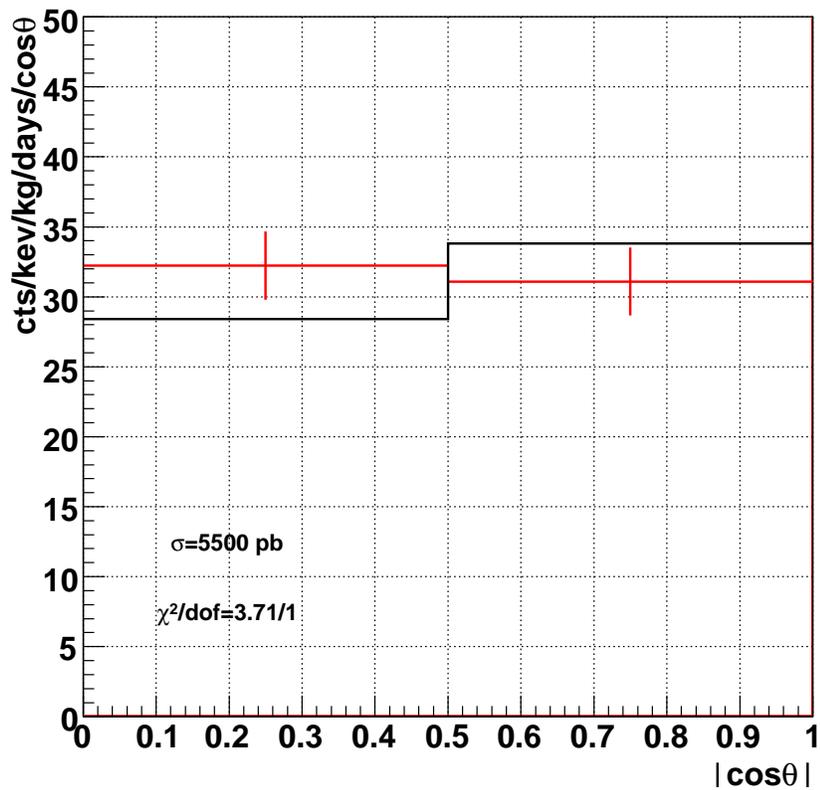}
\caption{\label{fig:costheta} 
Measured $|\cos \theta|$ distributions of Kamioka Run-5 
(histogram with error bars, red online)
and the best-fitted expected $|\cos \theta|$ distribution 
(histogram without error-bars, black online). 
The energy range and mass of the WIMP are  100--120 keV$\alpha$.e.
and 100 GeV/$\rm c^2$, respectively.
The best-fitted cross section of 5500 pb yielded $\chi^2$/dof = 3.71/1,  and  
this expected WIMP signal was rejected at a 90$\%$ confidence level 
by a $\chi^2$ test. 
}
\end{figure}

We finally set the direction-sensitive dark matter limits.
Here we conservatively treated all of the 1244 nuclear recoil events 
as dark matter events without any background subtraction.
Because the $|\cos \theta|$ distribution has an energy dependence and 
the statistics were not very large, we 
made two-bin $|\cos \theta|$ histograms for 15 energy ranges. 
We show one of the $|\cos \theta|$ histograms with a 
best-fitted expected WIMP signal in  FIG. \ref{fig:costheta}.
The energy range and mass of the WIMP are 100--120 keV$\alpha$.e. and 
100 GeV/$\rm c^2$, respectively.
The best-fitted cross section of 5500 pb yielded 
a $\chi^2$/dof of 3.71/1 and  
this expected WIMP signal was rejected at a 90$\%$ confidence level  
by a $\chi^2$ test in this case. 
We then fitted the $|\cos \theta|$ distributions with the expected WIMP signals 
for the other 14 energy ranges.
We took the smallest cross section 
as a limit for the WIMP mass (in this case, 100 GeV/$\rm c^2$).
We calculated the limits for WIMPs with masses from 
30 GeV/c$^2$  to 1000 GeV/c$^2$ in the same manner.
The WIMP-signal $|\cos \theta|$ distributions expected with all of the WIMP masses
were rejected  by $\chi^2$ tests. 

\begin{figure}
\includegraphics[width=1.\linewidth]{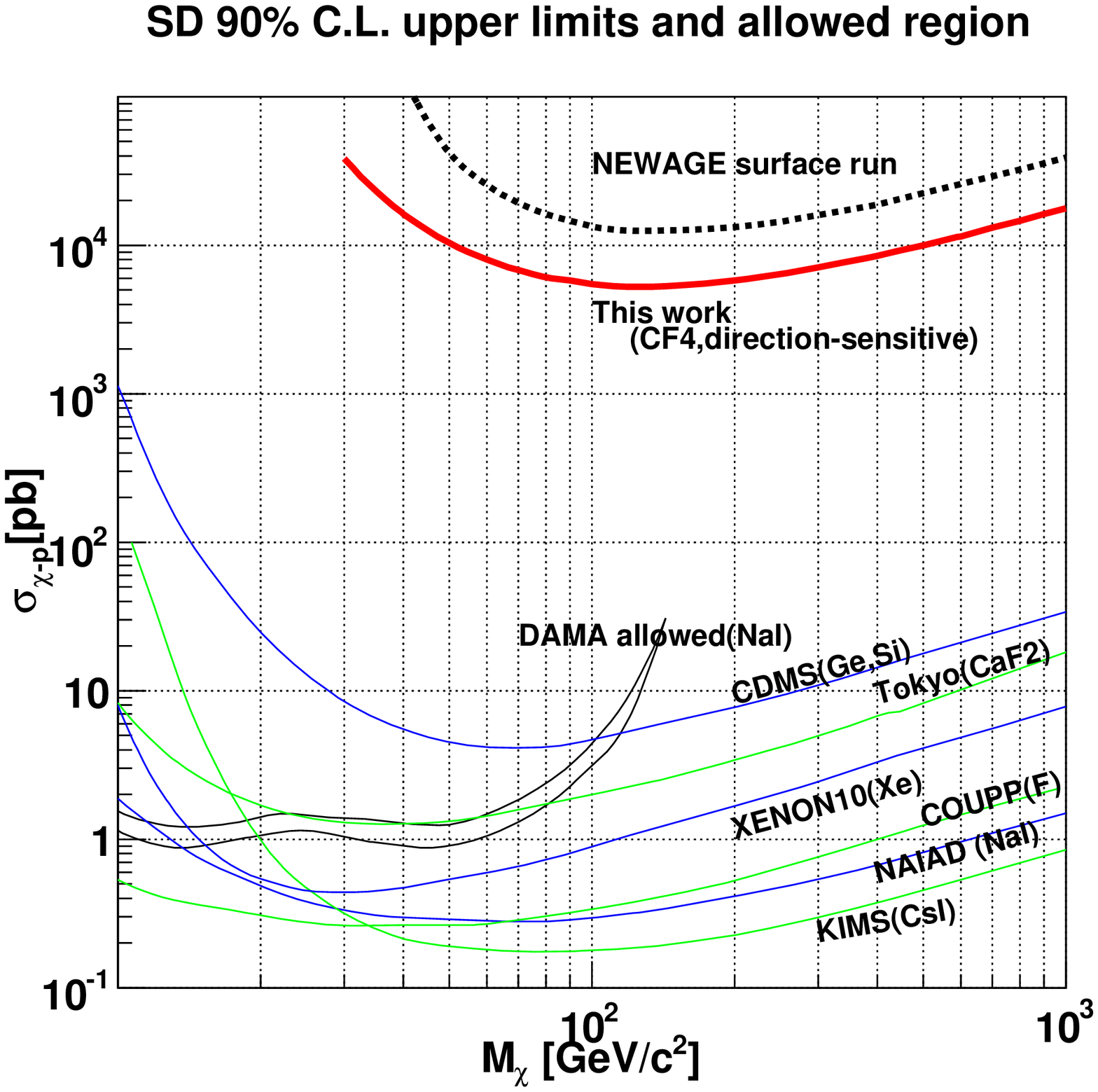}
\caption{\label{fig:limit} 
Upper limits and allowed region in the WIMP-proton spin-dependent 
cross section versus WIMP mass parameter space.
Thick solid line shows the limits obtained 
in this work.
Limits from our surface run\cite{ref:NEWAGE_PLB2007} are shown 
by a thick-dotted line for comparison.
Limits and allowed region 
from other direction-insensitive WIMP-search experiments 
(DAMA(NaI)\cite{ref:DAMA_annual2000}, CDMS\cite{ref:CDMS_SD}, 
Tokyo $\rm CaF_2$\cite{ref:Tokyo_CaF2}, XENON10\cite{ref:XENON10_SD},
COUPP\cite{ref:COUPP}, NAIAD\cite{ref:NAIAD}, 
KIMS\cite{ref:KIMS}) are shown for reference. }
\end{figure}

The obtained upper limits of the SD
WIMP-proton cross section are shown in FIG.~\ref{fig:limit} .
The best limit was 5400pb for WIMPs with a mass of 150 GeV/c$^2$. 
This result marked a new
sensitivity record for an SD WIMP search with the 
direction sensitive method.


We also fitted the
measured $|\cos \theta|$ distribution with the isotropic background model. 
The fitting result gave $\chi^2$/d.o.f. 
= 0.110/1 independent of the
WIMP mass, and the isotropic background model was not rejected at 90$\%$ 
confidence level. 

\begin{table}
  \caption{\label{tab:params}Astrophysical and nuclear parameters used to calculate the WIMP-proton cross section limits.}

  \begin{tabular}{ll}
    WIMP velocity distribution&Maxwellian\\\hline
    Solar velocity&$v_s=244 {\rm km s ^{-1}}$\\
    Maxwellian velocity dispersion&$v_0=220{\rm km s^{-1}}$\\
    Escape velocity&$v_{\rm esc}={\rm 650 km s^{-1}}$\\
    Local halo density&$0.3{\rm GeVcm^{-3}}$\\
    Spin factor of $^{19}$F&$\lambda^{2}J(J+1)=0.647$\\\hline

\end{tabular}
\end{table}

\section{Discussions}
\subsection{Background}
We studied the origin of the background for future 
improvement of the sensitivity.
We discuss the estimated contribution 
of various background sources to the count rate  
at the energy threshold of 100 keV$\alpha$.e.
Details will be reported elsewhere.

We started the study with the background from outside the vessel.
The measured flux of fast neutrons 
was 1.9$\rm \times$10$\rm ^{-6} cm^{-2}s^{-1}$ \cite{ref:minamino}, and  
their contribution at 100 keV$\alpha$.e. 
was simulated to be less than 0.2 counts/keV/kg/days.
We measured the $\gamma$-ray flux in our laboratory and simulated 
the contribution of  $\gamma$-rays, considering the 
$\gamma$-ray detection efficiency.
The contribution of $\gamma$-rays
was simulated to be less than 7$^{+7}_{-4}$ counts/keV/kg/days.
The cosmic-ray muon flux at Kamioka Observatory is 
$\rm 6 \times 10^{-8}cm^{-2}s^{-1}sr^{-1}$
\cite{ref:SK_NIM}.
The muon rate passing
through the 
effective volume of the $\mu$-TPC was calculated to be less than 
$2\times 10^{-4}$ counts/s.
We assumed the rejection power 
of electrons, $8\times 10^{-6}$,
as an upper limit to that of muons.
Then the rate of muon tracks 
misidentified as nuclear tracks 
should be less than
0.2  counts/kg/days above 100 keV$\alpha$.e.
The secondary particles produced by cosmic-ray
muons are studied as ambient gamma rays and neutrons in previous 
discussions.

Next, we studied background 
sources within the vessel. 
Radioactive isotopes in the $^{238}$U and $^{232}$Th chains are 
the main background sources in many rare-event measurements. 
We first studied the contribution of the radon gas 
($^{220}$Rn, $^{222}$Rn, and their progeny) 
by a radon-rich run.
(We intentionally contaminated the detector with 
radon gas 
in the same way as we unintentionally did at the beginning of Run5-2.)
We found that the radon gas accounted for
about 5 counts/keV/kg/days at 100keV$\alpha$.e.
We then studied the contribution of alpha particles emitted from 
materials exposed to the detection volume.
It is known that 
alpha particles from the surface of the materials 
would deposit part of the entire energy in the detection 
volume and would make background events in the DM energy range.
We simulated the energy depositions of 
these surface-$\alpha$ events and found that 
the upper limits of their contribution can be known by the 
measured count rate around 1 MeV$\alpha$.e.
We thus found that these surface-$\alpha$ events, 
other than the partial energy deposition in the thin (5 mm) volume 
between the $\mu$-PIC and the GEM (gap events, see FIG. \ref{fig:TPC}),  
made negligible contributions.
The gap events were found to potentially 
explain the remaining background.
We found that the relative gas gains of the GEM and $\mu$-PIC affect 
the contributions of gap events to the count rate around 100keV$\alpha$.e; 
namely, a higher GEM-gain yields a  lower count rate and vice versa.
Because higher GEM-gain operation would increase the risk of 
fatal damage to the GEM, 
we performed a lower GEM-gain run to test this possibility. 
The count rate actually increased in the lower GEM-gain run.


We summarize the results of the background study in Table~\ref{tab:BG}.
The results indicate we can potentially explain the background with the 
the gap events and ambient gamma-ray events. We are making efforts to 
reduce these two main sources of the background to improve the sensitivity.

\begin{table}
\caption{\label{tab:BG}
Results of background studies.
The rates at the energy threshold
are shown in the unit of [counts/kg/days/keV] .}
\begin{tabular}{ll}
source & rate\\
\hline

\hline
Ambient gammas&$\sim 10$\\
Ambient fast neutrons&$\sim10^{-1}$\\
Cosmic muons&$<2\times10^{-1}$\\
\hline
Internal $\alpha$(fiducial volume)&$<10^{-1}$\\
Internal $\alpha$(gap volume)&$<40$\\
Internal $\beta$&$<5$\\
\hline
Measured(Run5) &50\\
\hline
\end{tabular}
\end{table}

\section{Conclusions}
\label{section:conclusion}
A direction-sensitive dark matter search experiment in the Kamioka 
underground laboratory with the NEWAGE-0.3a detector was performed.
The measurements were performed from September 11, 2008 until December 4, 
2008,  
producing a total exposure of 0.524 kg$\cdot$days.
As a result of this experiment, 
improved spin-dependent WIMP-proton cross section limits 
by a direction-sensitive method,
including a new record of 5400 pb for 150 GeV/c${^2}$ WIMPs, 
were achieved.

\section*{Acknowledgments}
This work was partially supported by Grant-in-Aids for KAKENHI 
(19684005) of Young Scientist(A); JSPS Fellows; and Global COE
Program ``The Next Generation of Physics, Spun from Universality
and Emergence'' from the Ministry of Education, Culture, Sports, 
Science and Technology (MEXT) of Japan.

\bibliographystyle{elsarticle-num}
\bibliography{NEWAGE2009.bib}

\end{document}